\documentclass[fleqn,usenatbib]{mnras}

\usepackage{newtxtext,newtxmath}

\usepackage[T1]{fontenc}

\DeclareRobustCommand{\VAN}[3]{#2}
\let\VANthebibliography\thebibliography
\def\thebibliography{\DeclareRobustCommand{\VAN}[3]{##3}\VANthebibliography}

\usepackage{graphicx}	
\usepackage{amsmath}	
\usepackage{enumitem}
\usepackage{amssymb}	
\newcommand{\abareff}{\bar{\alpha}_{\mathrm{eff}}}

\title[Convection in CEs and the formation of DWDs]{Convection and Spin-Up During Common Envelope Evolution: \\ The Formation of Short-Period Double White Dwarfs}

\author[Wilson \& Nordhaus]{
E. C. Wilson$^{1}$\thanks{E-mail: ecw7497@rit.edu} and 
J. Nordhaus$^{1,2}$\thanks{E-mail:nordhaus@astro.rit.edu}
\\
$^{1}$Center for Computational Relativity and Gravitation, Rochester Institute of Technology, NY 14623, USA
\\ $^{2}$National Technical Institute for the Deaf, Rochester Institute of Technology, NY 14623, USA
}

\date{Accepted XXX. Received YYY; in original form ZZZ}

\pubyear{2020}

\begin{document}
\label{firstpage}
\pagerange{\pageref{firstpage}--\pageref{lastpage}}
\maketitle

\begin{abstract}
The formation channels and predicted populations of double-white dwarfs (DWDs) are important because a subset will evolve to be gravitational-wave sources and/or progenitors of Type Ia supernovae.  Given the observed population of short-period DWDs, we calculate the outcomes of common envelope evolution when convective effects are included. For each observed white dwarf in a DWD system, we identify all progenitor stars with an equivalent proto-WD core mass from a comprehensive suite of stellar evolution models. With the second observed white dwarf as the companion, we calculate the conditions under which convection can accommodate the energy released as the orbit decays, including (if necessary), how much the envelope must spin-up during the common envelope phase.  The predicted post-CE final separations closely track the observed DWD orbital parameter space, further strengthening the view that convection is a key ingredient in common envelope evolution.

\end{abstract}




\section{Introduction}
Common envelopes (CEs) are short, yet highly critical, phases in the evolution of binary systems \citep{Paczynski1976}. For two main-sequence stars with initial separations $\lesssim$5-7 AU, post-main-sequence expansion may result in a common envelope phase. As the primary star's radius expands during the giant phases, the orbit can destabilize.  Engulfment of the secondary by the primary's envelope can occur either directly, or by other processes such as orbital decay via tidal dissipation \citep[e.g.,][]{Nordhaus2010TidesDwarfs,Nordhaus2013, Ivanova2013,Kochanek2014,Chen2017}.  

Once immersed, the secondary and the primary's core orbit inside a common envelope.  The orbit decays rapidly as energy and angular momentum are transferred to the primary's envelope \citep{Iben1993,Nordhaus2006,Nordhaus2007}.  Two outcomes can occur: (i.) the primary's envelope is ejected leaving a short-period, post-CE binary or (ii.) the secondary is destroyed during the CE leaving a single star that had its evolution significantly modified by the secondary \citep{Nordhaus2011, Guidarelli2019}.

CE evolution is thought to be the primary, though not sole, mechanism for producing the Universe's short-period binaries \citep[$a\lesssim R_\odot$; e.g.,][]{Toonen2013,Canals2018,Kruckow2018,Fabrycky2007, Thompson2011, Shappee2013,Michaely2016}. Since the CE phase spans only a short fraction of the binary's lifetime, direct detection is difficult. For this reason, identification of CEs rests in their precursor emission \citep{MacLeod2018} and their progeny, e.g., short-period binaries and associated objects such as planetary nebulae \citep{Iben1993,Ivanova2013, Jones2020}.

A widely studied post-CE outcome is that of short-period double white dwarfs (DWDs)  \citep{Webbink1984,MarshEtal1995,Iben1990}.  DWDs are important binary systems as some are thought to be progenitors of Type Ia supernovae (SN Ia).  In addition, DWDs are potential strong mHz gravitational-wave sources and may be detectable by future missions such as \textit{LISA} \citep[e.g.,][]{Webbink1984, Iben1984, Ivanova2013,Marsh2011,Kilic2015, Brown2011, Ruiter2010}.

Observations of DWD candidates have allowed for lengthy compilations of their orbital parameters \citep{Saffer1988,Marsh1995, MarshEtal1995, Holberg1995, Moran1997a, Maxted2000,Maxted2002:DWD,Maxted2002:RV,Napiwotzki2002,Karl2003,Karl2003a,Nelemans2005,Morales-Rueda2005,Kilic2007,Badenes2009,Kilic2009,Mullally2009,Kilic2010,Kulkarni2010,Steinfadt2010,Vennes2011,Bours2014,Bours2015,Debes2015,Brown2016, Hallakoun2016,Santander-Garcia2016, Rebassa-Mansergas2017,Brown2020}. With these data, statistical studies, in combination with binary population synthesis (BPS) studies, have provided an at-large view of the DWD population, often noting that models over-predict the number of observable DWD systems \citep[e.g.,][]{Ferrario2012,Marsh2011,Maoz2018,Toonen2017}.

A subset of the observed DWD population is that of extremely low-mass (ELM) white dwarfs (WDs). Typical, single WDs have been observed to follow an initial-final mass relation (IFMR) which describes the correlation between initial stellar mass and final WD mass following a single star's full evolution. A semi-empirical IFMR has been determined for single star systems \citep{Cummings2018}. However, stars in binary systems and especially ELM WDs have masses which are too low when compared to the IFMR. The ``undermassive'' nature of these WDs is consistent with the evolution of the star having been interrupted by some binary interaction whereby the outer layers of the star have been stripped \citep{Webbink1984,MarshEtal1995,Iben1990}. The interruption of evolution can occur when the primary is on the Asymptotic Giant Branch (AGB) and the proto-WD core is nearly at its IFMR-predicted final mass or the interruption can occur when the primary is on the Red Giant Branch (RGB) and the proto-WD core is just a fraction of the IFMR-predicted final mass \citep{Kilic2007a}.

Inclusion of convective effects in CEs has been shown to produce M-dwarf+WD binaries with periods matching observations, a significant improvement over the results of binary population synthesis studies \citep{Wilson2019}. Convection transports the energy released by the shrinking orbit to the surface where it is radiated away.  This allows the M-dwarf to travel deeper into the primary unbinding the envelope, thereby producing sub-day periods consistent with observations. Convection is also important for the production of Type Ia supernovae as it allows transport of energy and angular momentum outward \citep{Soker2013}.

In this work, we investigate the effect that convection in common envelopes has on the formation of double white dwarfs. Our initial conditions consist of a white dwarf companion that enters a CE with an evolved star.  We calculate the conditions under which convection can accommodate the energy released as the orbit decays including (if necessary) how much the envelope must spin-up during the CE.  Under these conditions, we compare the predicted outcomes of common envelope evolution to the largest observational sample of DWDs to date.

In Section~\ref{sec:Wilson2019}, we describe how convection in conjunction with radiative losses from the surface effect the outcomes of CE evolution. In Section~\ref{sec:Methods} we describe the observational sample of DWDs, the stellar models employed, and the physics included in our models. Sections~\ref{sec:results} and \ref{sec:discussion} present the results of our analysis as well as a physical interpretation of the theory and a comparison to the observations. We conclude in Section~\ref{sec:conclusion} and comment on future directions.

\section{How convection impacts common envelope evolution}
\label{sec:Wilson2019}

Global simulation studies have focused on determining the necessary physical processes required to successfully eject the CE.  However, these studies have typically neglected effects such as convection and radiation due to computational complexity. Instead, many explore additional energy sources (e.g., recombination, accretion, jets) and longer-term processes  \citep{Ricker2008, Ricker2012, Ivanova2015,Nandez2015,Soker2015,Kuruwita2016, Sabach2017,Glanz2018,Grichener2018,Ivanova2018,Kashi2018,Soker2018,Reichardt2020}. However, RGB and AGB stars possess deep and vigorous convective envelopes making convection a necessary ingredient for the physical fidelity of CE simulations.

In lieu of simulations, a widely-used, back-of-the-envelope energy argument for estimating how efficiently two stars exchange energy during a CE is often characterized by a constant value, $\abareff$, typically defined as:
\begin{equation}
	\bar{\alpha}_{\rm eff}=\frac{E_{\rm bind}}{\Delta E_{\rm orb}},
\end{equation}
where $E_{\rm bind}$ is the binding energy of the envelope and $\Delta E_{\rm orb}$ is the change in the companion's orbital energy due to inspiral \citep[e.g.,][]{Tutukov1979, Iben1984,Webbink1984,Livio1988,DeMarco2011, Iaconi2019}. This value is often taken to be a constant, though in principle, it should functionally depend on the binary parameters and internal structure of the CE.  For example, the location and depths of the convective zones were shown to greatly impact where energy can be tapped to drive envelope ejection \citep{Wilson2019}.

Population synthesis studies which use an $\alpha$-prescription that is not dependent on the internal structure nor the age of the primary find that very low efficiencies best reproduce observations. These same studies over-produce longer-period binaries even though short-period binaries are readily observed in nature \citep{Politano2007,Davis2010,Zorotovic2010, Toonen2017}. For DWDs, many studies demonstrate that use of the standard $\alpha$-prescription does not match observations. In fact, in some cases, the standard $\alpha$-prescription requires unphysical efficiencies in order to form DWDs \citep{VanDerSluys2006,Woods2011}.

An $\alpha$-prescription which is physically motivated and dependent on the interior structure of the star may better explain the ejection efficiency within CEs. In CE systems, energy released as the orbit decays can be transported via convection to an optically-thin layer of the primary where it is radiated away \citep{MacLeod2018a,Wilson2019}.  To model convection in CEs, the convective transport timescale of the primary is compared to the inspiral timescale. Where the convective transport timescale is short compared to the inspiral timescale, convection can carry orbital energy to the surface where it is lost. In this case, energy would not contribute to unbinding the envelope and thus, $\alpha_{\rm eff}$ would be consistent with zero in these regions. In the opposite case, where the inspiral timescale dominates, energy from the decaying orbit can only be used to raise the negative binding energy of the envelope; where this occurs, $\alpha_{\rm eff}=1$. The convective transport timescale, inspiral timescale, and ejection efficiencies are all functions of the radial position, $r$, inside the primary. The ejection efficiency for the CE phase can then be found by averaging radially-dependent efficiency values in the following way:

\begin{equation}
\abareff = \frac{\int_{r_{\rm{i}}}^{r_{\rm{f}}}\alpha_{\rm eff}[r] \rm{d}E_{\rm{orb}}[r]}{E_{\rm{orb}}[r_{\rm{f}}]-E_{\rm{orb}}[r_{\rm{i}}]}.
\label{eq:abar}
\end{equation}

When this physically motivated $\alpha$-prescription is used, the ejection efficiency and final orbital period are dependent on the size of the surface-contact convective region (SCCR) within the primary. Since the companion can typically travel through this convective region without contributing any energy to unbinding the envelope, a larger SCCR corresponds to a shorter final orbital period, as the companion travels closer to the primary's core before energy can be tapped to drive ejection. The final separation is therefore related to the evolutionary stage of the primary, as the SCCR depth fluctuates with stellar age.  In this work, we investigate how the inclusion of convection in common envelope evolution effects the emergent population of DWDs.

\section{Methods}
\label{sec:Methods}
With a compilation of observed DWDs and their orbital parameters, as well as stellar interior models, we compare the population of DWDs to modelled CE outcomes. A graphical summary of our method is portrayed in Figure~\ref{fig:MethodCartoon}, with panels \textbf{A}, \textbf{B}, and \textbf{C} corresponding to Sections~\ref{sec:MethodObs}, \ref{sec:MethodModels}, \ref{sec:MethodConvCE}, respectively.

\subsection{Observations}
\label{sec:MethodObs}
To date, many DWDs have been observed and characterized primarily via radial velocity and/or transit methods \citep{Saffer1988,Marsh1995, MarshEtal1995, Holberg1995, Moran1997a, Maxted2000,Maxted2002:DWD,Maxted2002:RV,Napiwotzki2002,Karl2003,Karl2003a,Morales-Rueda2005,Nelemans2005,Kilic2007,Badenes2009,Kilic2009,Mullally2009,Kilic2010,Kulkarni2010,Steinfadt2010,Vennes2011,Bours2014,Bours2015,Debes2015,Brown2016, Hallakoun2016,Santander-Garcia2016, Rebassa-Mansergas2017,Brown2020}. Our sample consists of 141 DWD systems each of which contain observed masses, the corresponding orbital periods, and statistical constraints on the secondary masses. We determine a separation for each system assuming the orbits are circular (Figure~\ref{fig:MethodCartoon}, panel \textbf{A}).

\subsection{Stellar Models}
\label{sec:MethodModels}
We use Modules for Experiments in Stellar Astrophysics (\textsc{MESA}, release 10108), an open-source stellar evolution code, to produce detailed, spherically-symmetric stellar interior models \citep{Paxton2010, Paxton2018}\footnote{\textsc{MESA} is available at http://mesa.sourceforge.net}.  The full evolution of the star is calculated for zero-age-main-sequence masses from $1.0M_\odot$ to $6.0M_\odot$ in increments of $0.2M_\odot$ with solar metallicity ($Z=0.02$). To match the semi-empirical initial-final mass relationship (IFMR) of \citet{Cummings2018}, we adopt a Reimers mass-loss prescription with $\eta_R=0.7$ on the RGB and a Bloecker mass-loss prescription with $\eta_B=0.15$ on the AGB \citep{Reimers1975, Bloecker1995}. Given these mass-loss coefficients, our evolutionary models match the observationally-derived IFMR within the measured errors \citep{Cummings2018}.

\begin{figure}
    \centering
    \includegraphics[width=0.9\columnwidth]{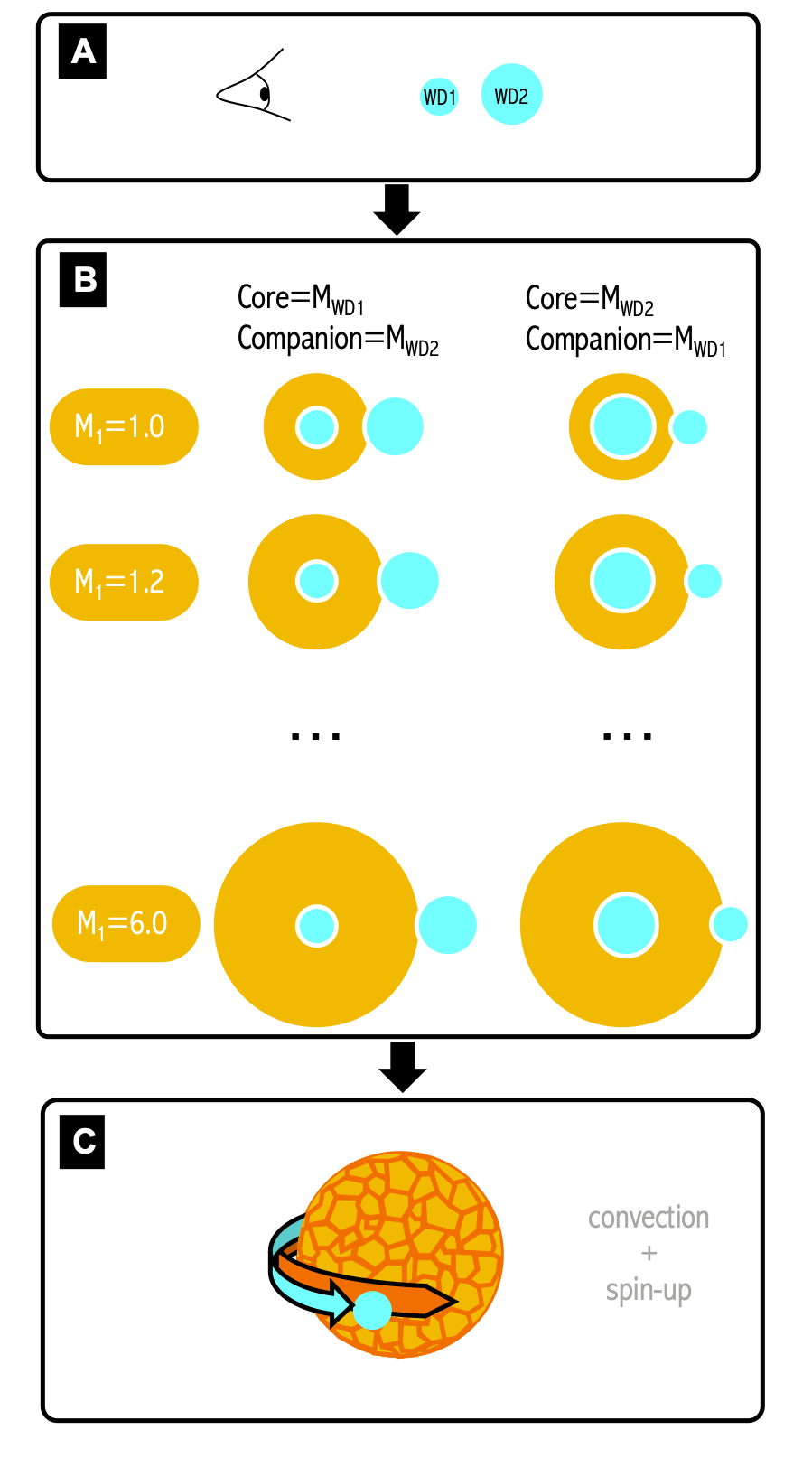}
    \caption{A cartoon of the method employed in this paper, in three panels. \textbf{A}: A list of DWD observations is used, which include WD masses and periods. Separations are estimated assuming circular orbits. \textbf{B}: Using stellar evolution models, we match the modelled core mass and the modelled CE companion to both of the observed/derived values for stellar models of primaries ranging from $1.0-6.0M_\odot$. \textbf{C}: Common envelope evolution includes convective effects as in \citet{Wilson2019} and spin-up of the envelope.}
    \label{fig:MethodCartoon}
\end{figure}

\begin{figure}
    \centering
    \includegraphics[width=\columnwidth]{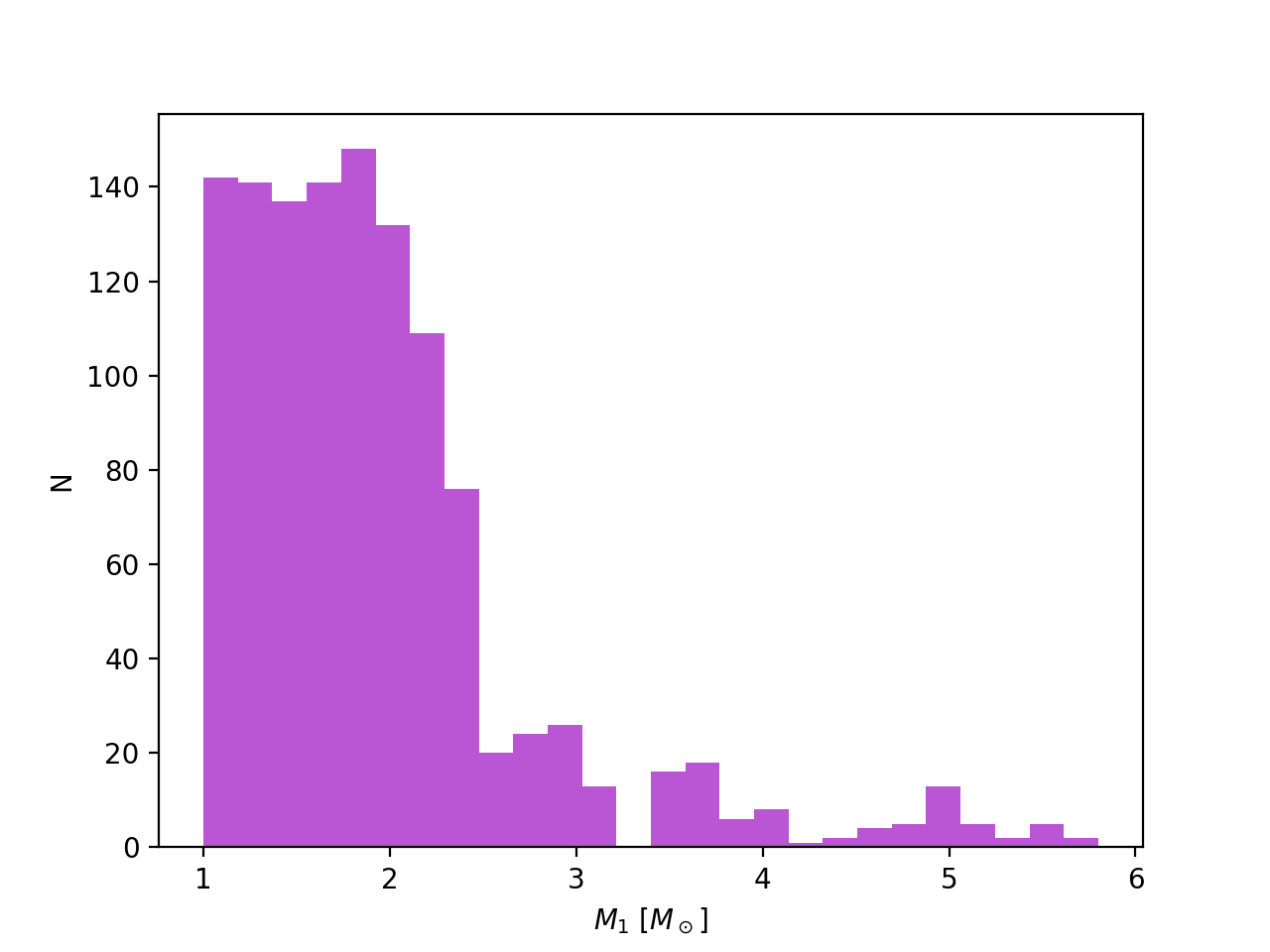}
    \caption{The frequency of primary mass values from which the cores are modelled. No initial mass function (IMF) is assumed from which to draw our data; this is effectively an IMF from which our data are drawn. Models were selected from a suite of primary masses ($1.0-6.0 M_\odot$) at a time in each primary's evolution when the mass of the core equals the mass of an observed double white dwarf component within 0.02$M_\odot$.}
    \label{fig:M1hist}
\end{figure}

For each WD mass in our observational sample of DWDs, we determine the time in each modelled star's evolution at which the core mass matches the observed WD mass to within $0.02M_\odot$ from a suite of primary mass models, ranging from $1.0-6.0M_\odot$.  This is depicted in Figure~\ref{fig:MethodCartoon}, panel \textbf{B}. With several initial mass primary models mapped to each observed system, the radius of the convective boundary of each modelled primary was found for this set of initial masses. We do not use any formal initial mass function (IMF).  Rather, we draw from all masses in our range, equally. This approach generates an initial mass distribution from the primaries which produced cores that match DWD observations. This distribution is shown in Figure~\ref{fig:M1hist}. We note that while the core mass for each model monotonically increases in time, several, higher-initial-mass ($M \ge 2.6M_\odot$) \textsc{MESA} models display a sharp, step-like discontinuity in core mass. In these cases, the core mass jumps from 0 to $\sim$8-10\% of the primary mass in a single timestep, preventing a match to observations for WDs with masses less than $\sim$8-10\% of the initial mass primary. In particular, the observed ELM WDs can only be matched by \textsc{MESA} models that exhibit continuous core growth (i.e. $M<2.6 M_\odot$).

\subsection{Modelling the CE with Convection and Spin-Up}
\label{sec:MethodConvCE}
For each observed DWD system, two corresponding CEs were modelled: (i.) the more-massive WD as the companion and the less-massive WD as the core mass, and (ii.) the less-massive WD as the companion and the more-massive WD as the core mass\footnote{We take the mass of the WD and the mass of the primary's core to be constant during common envelope evolution.  The accretion rate onto a WD at the Eddington limit is given by $\dot{M}_{\rm Edd}\simeq3\times10^{-5}\left(R/R_{\rm WD}\right)M_\odot yr^{-1}$.  Even if the CE were to last $10^3$ years, accretion would only increase the WD's mass by three percent.}. These two modelled systems were then iterated through several initial-mass primaries as shown in Figure~\ref{fig:MethodCartoon}, panel \textbf{B}.

Convective effects of the primary were taken into account by comparing the inspiral timescale,
    \begin{equation}
	t_{\mathrm{inspiral}}[r] = \int_{r_{\mathrm{i}}}^{r_{\mathrm{shred}}}{\frac{\left(\frac{\mathrm{d}M}{\mathrm{d}r}-\frac{M[r]}{r}\right)\ \sqrt[]	{v_r^2+(\bar{v}_{\phi}[r]^2+c_s[r]^2)^2}}{4 \xi \pi G m_{\rm comp} r \rho[r]}\mathrm{d}r },
	\label{eq:tinsp}
\end{equation}
   \citep{Nordhaus2006} to the convective transport timescale,
\begin{equation}
	t_{\mathrm{conv}}[r] = \int_{r_{\rm i}}^{R_{\star}}\frac{1}{v_{\mathrm{conv}}[r]} \mathrm{d}r,
	\label{eq:tconv}
\end{equation}
\citep{Grichener2018} where $r_i$ is the initial radial position and $r_{\rm shred}$ is the tidal shredding radius of the companion, approximated as $r_{\mathrm{shred}}\simeq R_{\rm comp}\sqrt[3]{2M_{\mathrm{core}}/m_{\rm comp}}$ \citep{Nordhaus2007}, where $R_2=r_{\rm WD}$, estimated via the WD mass-radius relation \citep{Chandrasekhar1933,Hamada1961, Wood1990,Wood1995}. The velocity terms $v_r$, $c_s$, $v_\phi$, $v_{\rm env}$, and $v_{\rm conv}$ are the radial velocity of the companion, the sound speed, the Keplerian velocity, envelope velocity, and the convective velocity, respectively. The relative velocity of the companion to the velocity of the envelope is given by $\bar{v}_\phi=v_{\phi}-v_{\rm env}$. We use $\xi=4$ to account for the geometry of the secondary's wake within the primary \citep{Park2017}. M is the enclosed stellar mass and $\rho$ denotes the primary's density. Terms which are shown with an r in square brackets are radially dependent.
    
Where the convective transport timescale is less than the inspiral timescale, energy liberated from the decaying orbit can be carried to the surface via convective eddies where it is lost from the system via radiation.  Note that in this regime, we assume the primary's radius does not appreciably expand and thus the liberated orbital energy does not contribute to unbinding the CE. For regions where the convective transport timescale is long compared to the inspiral timescale, energy must be deposited locally in the gas and thus can only be used to raise the negative binding energy of the envelope.  The orbit continues to shrink until the envelope is either unbound, leaving a post-CE DWD, or the companion tidally disrupts inside the CE, leaving a single star whose evolution has been significantly modified.

Convection in CEs is an important physical effect to investigate as RGB and AGB stars have deep and vigorous convective envelopes. In addition to transporting energy, convection also acts to distribute energy throughout the envelope. While the depth of the convective region changes on hundred-year timescales, the inspiral timescales are short (often $\lesssim$1 year, but at most $\sim$30 years).  Thus, we perform an analysis of the effects of convection for a single snapshot in the primary's evolution.

In addition to convection, we also consider spin-up of the envelope during common envelope evolution as simulations of CEs have shown significant transfer of angular momentum from the orbit to the gas \citep{Ricker2012,MacLeod2018,Chamandy2018}. As RGB/AGB stars are slow rotators, we assume that each primary is initially stationary.  As the companion inspirals through the primary, the envelope velocity, $v_{\rm env}$, can increase as it begins to spin until it reaches co-rotation, where $v_{\rm env}=v_\phi$ and orbital decay is halted. In order for convection to transport the companion's orbital energy, the maximum luminosity that subsonic convection can accommodate,
\begin{equation}
    L_{\rm conv, max}[r]= 4 \pi \rho[r] r^2 c_s[r]^3,
    \end{equation}
    \citep{Quataert2012, Shiode2014, Sabach2017} must be greater than the drag luminosity,
    \begin{equation}
    L_{\rm drag}[r]=\xi \pi R_{\rm acc}^2 \rho[r](v_{\phi}[r]-v_{\rm env})^3,
    \label{eq:Ldrag}
    \end{equation}
    where $R_{\rm acc}$, the accretion radius, is given by
    \begin{equation}
    R_{\rm acc}~=~\frac{2 G m_{\rm comp}}{(v_{\phi}[r]-v_{\rm env})^2+c_s[r]^2},
    \end{equation}
\citep{Nordhaus2006}.
If the relative velocity between the orbit, $v_{\phi}$, and the envelope, $v_{\rm env}$, is reduced, the $L_{\rm drag} \le L_{\rm conv,max}$ constraint is more readily met since the inspiral timescale increases. The inspiral timescale may be increased such that the convective transport timescale becomes dominant, thus allowing the companion to travel deeper into the primary before contributing energy to unbind the envelope, thereby decreasing the ejection efficiency. When $L_{\rm drag} \le L_{\rm conv, max}$, the nature of convection is unchanged; if the opposite is true ($L_{\rm drag} > L_{\rm conv, max}$) convection will transition to the supersonic regime where orbital energy can converted to kinetic energy via shocks, thereby making the primary's envelope less bound. 
    
For the orbital energy released during inspiral through the convective zone to be fully transported and radiated away, some amount of spin-up of the envelope may be required to satisfy the luminosity condition.  The amount required is calculated by first representing the relative velocity of the two bodies as a fractional velocity: $v_{\phi}[r]-v_{\rm env}=\beta v_{\phi}[r]$ which is then substituted into Equation~\ref{eq:Ldrag}. The drag luminosity and the maximum luminosity that convection can accommodate are then equated, i.e., $L_{\rm drag} = L_{\rm conv, max}$. By setting $r=r_{\rm SCCR}$, the base of the surface-contact convective region\footnote{The surface-contact convective zone (SCCR) radius, $r_{\rm SCCR}$, is defined as the radius from the center of the primary to the base of the convective region which extends to the stellar surface.} the solutions for $\beta$ can be found by solving the following fourth-order equation:
\begin{equation}
    v_{\phi}[r]^4 \beta^4 - \frac{\xi G^2 m_{\rm comp}^2 v_{\phi}[r]^3}{r^2 c_s[r]^3}\beta^3 + 2v_{\phi}[r]^2c_s[r]^2\beta^2 + c_s[r]^4 = 0.
	\label{eq:beta4root}
\end{equation}
Note that because this equation is evaluated $r=r_{\rm SCCR}$, the solutions are the spin-up values at the base of the convective zone; if the region spanning from the base of the convective zone to the surface were rotating at this value, the luminosity condition, $L_{\rm drag} \le L_{\rm conv, max}$ would easily be met.

For this analysis, only real roots of the above equation are considered. For systems with $\beta=0$, the companion and the stellar envelope are fully co-rotating. For systems with $\beta=1$, the stellar envelope is stationary. We assume that the envelope is initially stationary, requiring the envelope to spin-up from stationary ($\beta=1$) to some velocity ($\beta \rightarrow 0 $) during the companion's inspiral; the real $\beta$ value closest to unity without exceeding it is used as the solution for each given system. The fraction of Keplerian speed taken on by the envelope is represented by $1-\beta$ (e.g., $\beta=0.7$ has an envelope spun-up to 30\% co-rotation) which is equivalent to $v_{\rm env}/v_\phi$.

To determine if the companion has deposited sufficient energy to unbind the envelope, we compare the primary's binding energy, $E_{\rm bind}$, to the companion's change in orbital energy, $\Delta E_{\mathrm{orb}}$, with an efficiency, $\abareff$, where $\abareff$ can be calculated via Equation~\ref{eq:abar}. We consider the envelope to be ejected, and therefore the final orbital separation of the pair to be, where the change in orbital energy is equal to the binding energy.

\section{Post-CE Orbital Separations}
\label{sec:results}
\subsection{Convection Alone}

During CE evolution, convection can transport the released orbital energy of the companion to the primary's optically-thin surface where the energy can be radiated away \citep{Soker2013,MacLeod2018a, Wilson2019}. This allows the companion to inspiral deep into the primary before reaching a region where convection can no longer sufficiently transport energy to the surface. Once the companion reaches this region, orbital energy cannot be radiated away and must contribute unbinding the envelope. This often occurs at the base of the convective zone where the inspiral timescale is greater than the convective transport timescale. The orbital energy liberated at the base of the convective zone is greater than the binding energy for many systems, and thus the envelope is often ejected here.

In many cases, the inclusion of convection predicts the post-CE binary's final separation to be the distance between the primary's core and convective boundary, $r_{\rm SCCR}$, and is consistent with the sub-day periods of post-CE, M-dwarf+WD pairs \citep{Wilson2019}. Given the observed population of short-period DWDs, we compare $r_{\rm SCCR}$ to the observed DWD separations, $a_{\rm obs}$. Because the same core mass can be produced via many different primary mass stars, each system cannot be mapped directly to a single model. Instead, for this comparison, each WD in the pair is reported with its $a_{\rm obs}$ separately and is compared to the population of stellar models with the same core mass. Across all observed DWD masses, we find a correlation between $a_{\rm obs}$ and $r_{\rm SCCR}$, underscoring a potential relationship between short-period binaries and convective properties of the primary. 

\begin{figure}
	\includegraphics[width=\columnwidth]{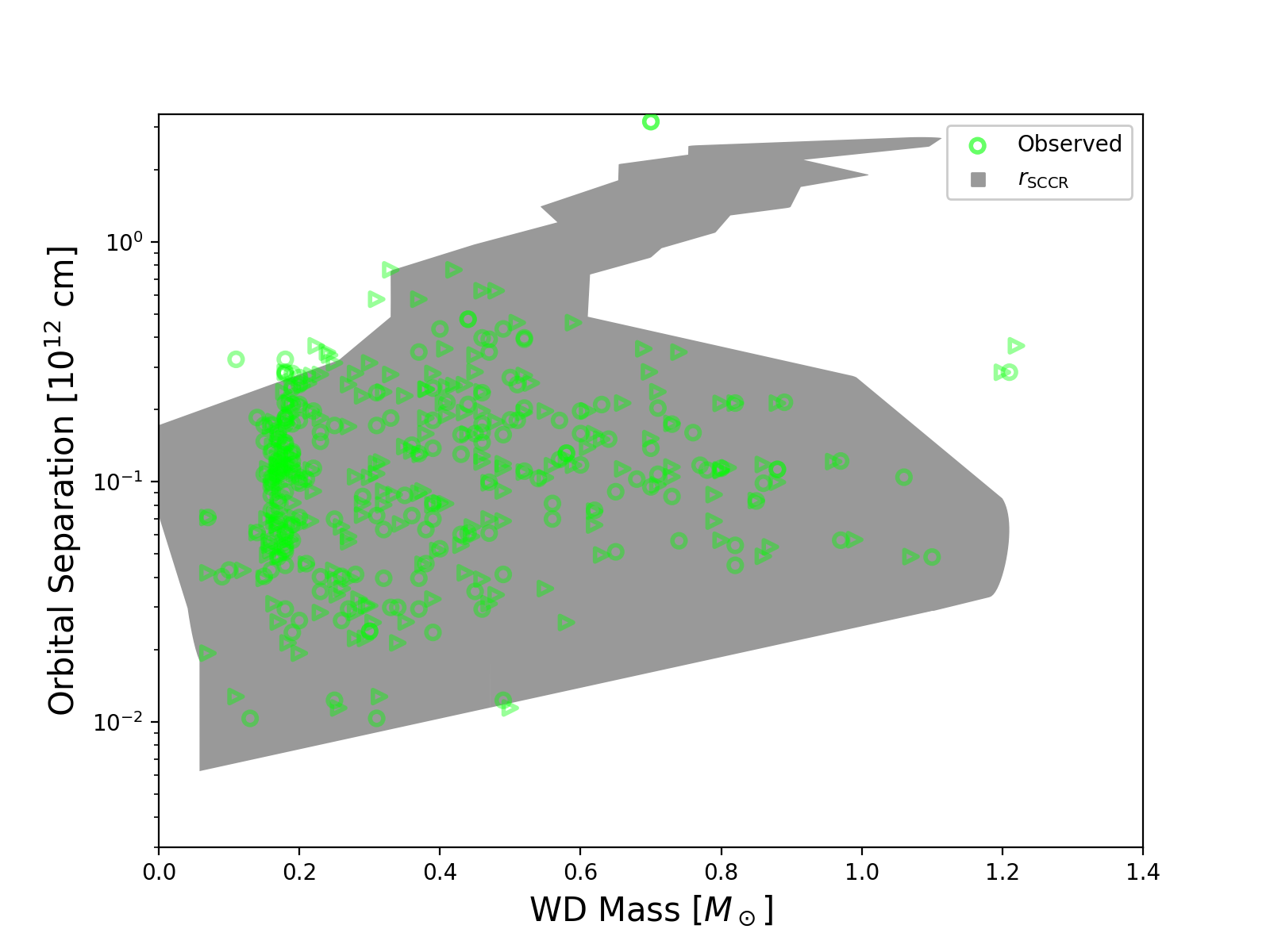}
    \caption{Comparison of observed DWD orbital parameters (green circles) and parameter space filled by the distance from the stellar core to the convective boundary (shaded grey). Each component mass of the DWD system is plotted individually. The parameter space filled by the models closely resembles the parameter space filled by the observed data. This correlation highlights the strong connection between convection in CEs and the post-CE DWD population.}
    \label{fig:volcano}
\end{figure}

The correlation between $a_{\rm obs}$ and $r_{\rm SCCR}$ can be seen in Figure~\ref{fig:volcano}. The 282 green points represent each white dwarf in the set of observed DWDs (141 pairs).  The grey shaded region is the space filled by the $r_{\rm SCCR}$ values of models with the same core mass as the observed DWD components. This novel correlation may indicate that WDs tend to halt their inspiral  (i.e., eject the envelope) shortly after exiting the convective zone. To determine under what conditions the engulfed WD would eject the envelope of the primary at the base of the convective zone, we included spin-up in our CE model which we discuss below.
   
\subsection{Convection and Spin-Up}
\label{sec:Result-ConvSpin}
In order to eject the envelope at the base of the convective zone, the following criteria must be met: (i.) the orbital energy released as the companion exits the convective zone must be in excess of the binding energy and (ii.) the maximum luminosity that convection can accommodate must be greater than the drag luminosity throughout the SCCR.  Since angular momentum is also transferred from the orbit to the gas, any spin-up of the envelope will lengthen the inspiral timescale.  This in turn relaxes the conditions for convection to transport energy to the surface.  

Since there is a limit to how much energy can be transported by subsonic convection, we calculate the amount of spin-up necessary to ensure that $L_{\rm drag} \le L_{\rm conv, max}$. At the SCCR, we allow the envelope to spin-up to some fraction of the Keplerian speed as determined by Equation~\ref{eq:beta4root}. While this equation yields four solutions, we choose the solution closest to unity without exceeding it.  When $\beta=1$, the envelope is stationary and thus as the envelope spins up, $\beta$ decreases until the envelope is in co-rotation with the orbit, i.e. $\beta=0$.  The value $1-\beta$ is equivalent to the ratio $v_{\rm env}/v_\phi$.

\begin{figure}
	\includegraphics[width=\columnwidth]{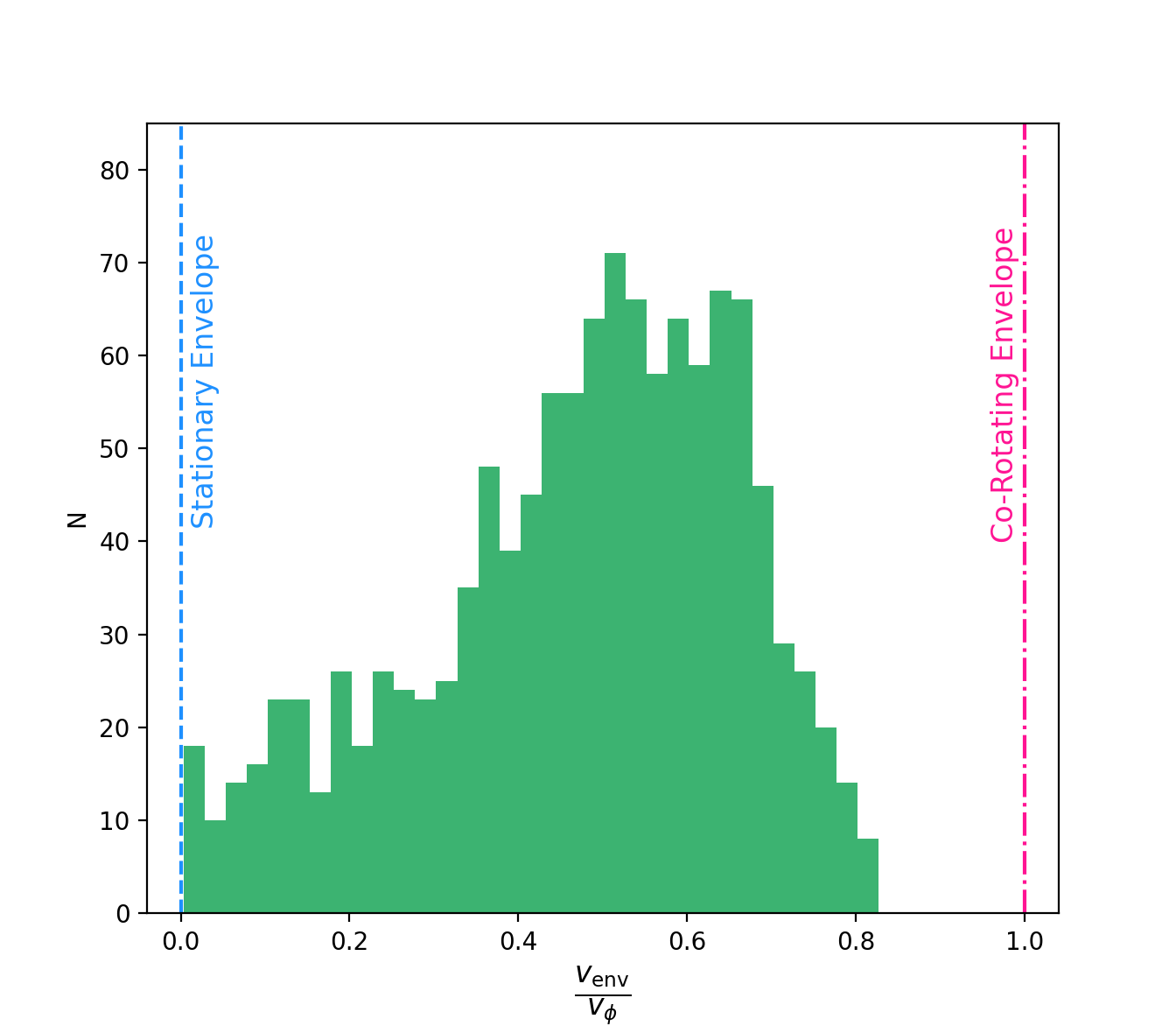}
    \caption{Frequency of $v_{\rm env}/v_\phi$ values in the total population of initial mass models. The $v_{\rm env}/v_\phi$ value is the fraction of co-rotation necessary for the inspiral timescale to increase enough for the maximum convective luminosity to exceed the drag luminosity, thus enabling convection to carry the energy to the surface. It falls off sharply after a peak at $v_{\rm env}/v_\phi=0.7$, indicating that an envelope spin-up of ${>}70\%$ of Keplerian is not commonly necessary.}
    \label{fig:betahist}
\end{figure}

For all but two modelled cases of our observed systems, there is a real, physical ($0<\beta<1$) solution.  However, we note that there are an additional two systems that lack a matching \textsc{MESA} model altogether. The frequency of $v_{\rm env}/v_\phi$ values peaks between $0.5$ and $0.7$, before sharply dropping off. There are very few systems with $v_{\rm env}/v_\phi>0.8$, indicating that it is rarely necessary for the envelope to reach 80\% of the Keplerian speed. A histogram of $v_{\rm env}/v_\phi$ values for the simulated initial systems is seen in Figure~\ref{fig:betahist}.

\subsection{Unbinding the Envelope}
\begin{figure}
    \centering
    \includegraphics[width=\columnwidth]{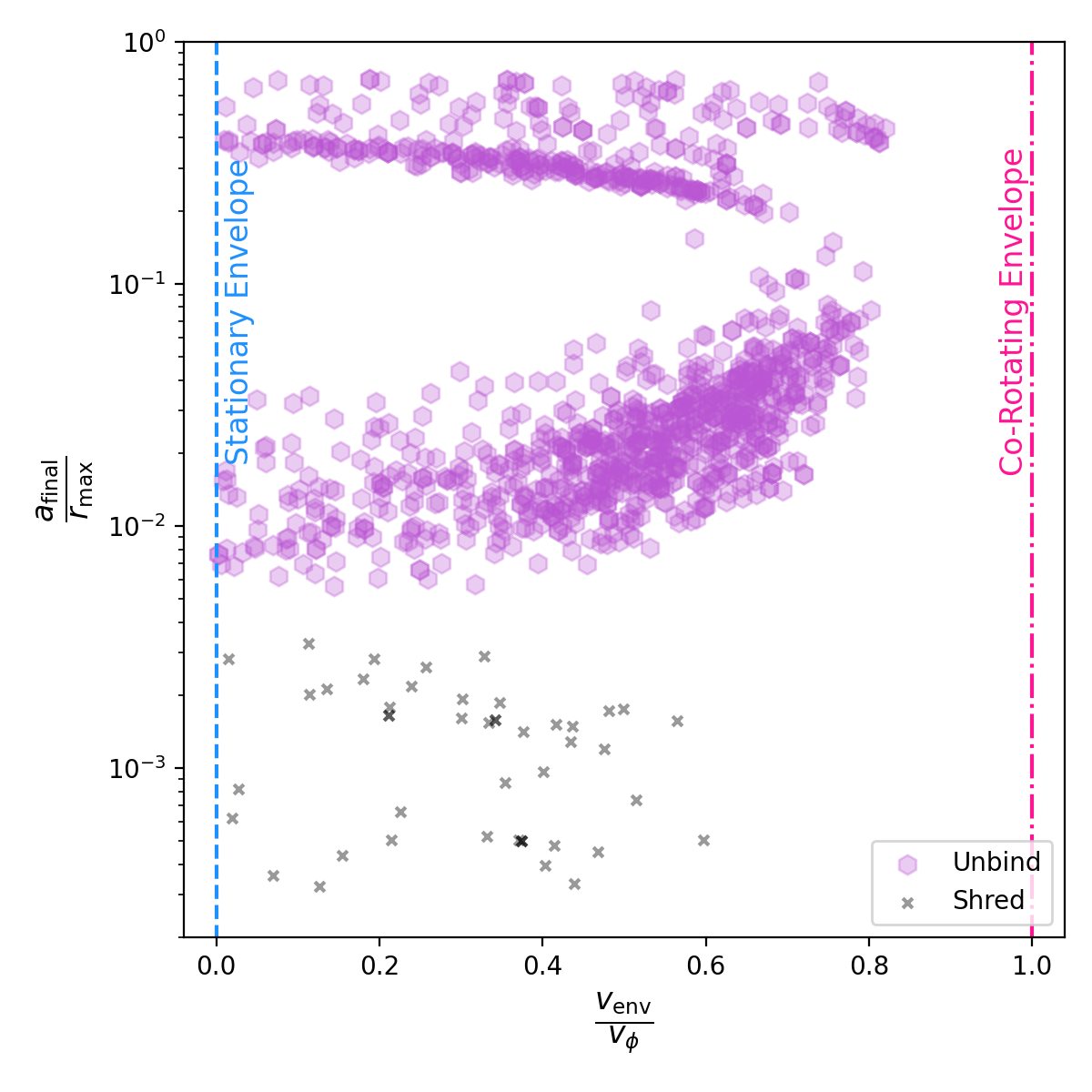}
    \caption{Final system orbital separation normalized to the radius of the primary with spin-up. Purple hexagons show systems where the companion has enough energy to unbind the envelope. Black Xs show systems where the companion shreds within the envelope. Most systems which shred and approximately half of systems that unbind have unphysical solutions for $\beta$ and are not plotted within these bounds.}
    \label{fig:unbindorshred}
\end{figure}

Though we do not formally draw from an IMF, we note how many post-CE, short-period binaries emerge from our initial mass pairings. The initial-mass primaries coupled with white dwarf companions are modelled via a convective CE with an envelope spun-up to their calculated $v_{\rm env}/v_\phi$. Of all modelled systems, 78\% have sufficient energy from the inspiraling companion to unbind the primary's envelope; the remainder tidally disrupt within the primary leaving binary-modified, single stars. In Figure~\ref{fig:unbindorshred}, the final orbital separation is shown, normalized to the radius of each initial mass primary, in purple hexagons. The systems which tidally shred are marked at their shredding radii with black Xs. A smaller $a_{\rm final}/r_{\rm max}$ corresponds to a shorter orbital period as the companion has traveled deeper into the envelope of the primary. The spin-up values for each system are calculated via Equation~\ref{eq:beta4root}.

\section{Discussion}
\label{sec:discussion}
\subsection{Convective, Spin-Up Model Matches Observations}

The correlation in parameter space displayed between the observed orbital separation vs. observed WD mass and the modelled convective boundary vs. modelled core mass (see Figure~\ref{fig:volcano}) is consistent with the ejection efficiency theory described in \citet{Wilson2019}. There is a degeneracy in initial mass of the modelled cores, and thus, a closer relationship cannot be determined. A given core mass can be modelled by up to 26 initial primary masses ($1.0-6.0 M_\odot$). Since core masses grow at varying rates for different primary masses, the state of the star when its evolution is interrupted, and therefore the location of the convective boundary, depends on the mass of the primary. The depth of the convective zone can also change on timescales on the order of ${\sim}10^2$ years for a single star. Though the modelled core masses match the observed core masses within a few percent, this variability combined with the lack of an IMF make this correlation intriguing but require future study for more robust conclusions to be drawn.

    \begin{figure}
    \centering
    \includegraphics[width=\columnwidth]{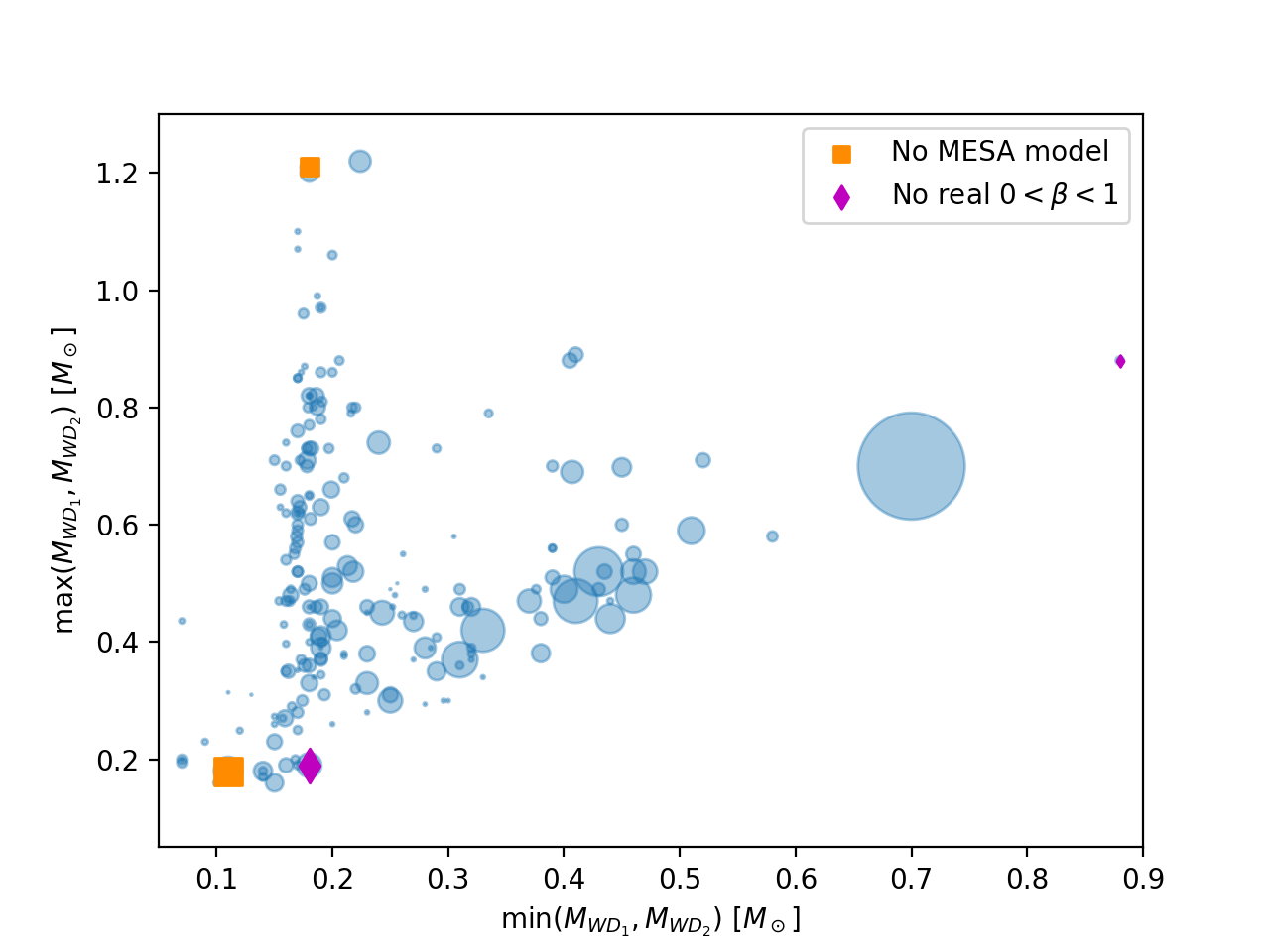}
    \caption{Mass-mass of each observed DWD system; marker size corresponds to orbital period. The orange squares mark the two systems that were not modelled due to lack of \textsc{MESA} models with core masses that matched observed WD masses. The magenta diamonds mark the two systems with only imaginary $\beta$ values. All four systems without a solution are on the periphery of parameter space.}
    \label{fig:sysNobeta}
\end{figure}

As described in Section~\ref{sec:Result-ConvSpin}, there is a solution for every initial mass system that was modelled\footnote{Note, two systems were not modelled as no \textsc{MESA} core was able to match the mass of the observed white dwarf to within $0.02M_\odot$.} with only two exceptions. This means that for each observed DWD system, there is a reasonable scenario where convection and spin-up will allow the envelope to be ejected at the base of the convective zone. Figure~\ref{fig:sysNobeta} displays all of the observed DWD systems in mass-mass space as well as the two systems that do not have a $\beta$ solution, and the two systems that do not have a corresponding \textsc{MESA} model. All four lie on the periphery.

\begin{figure}
    \centering
    \includegraphics[width=1.15\columnwidth]{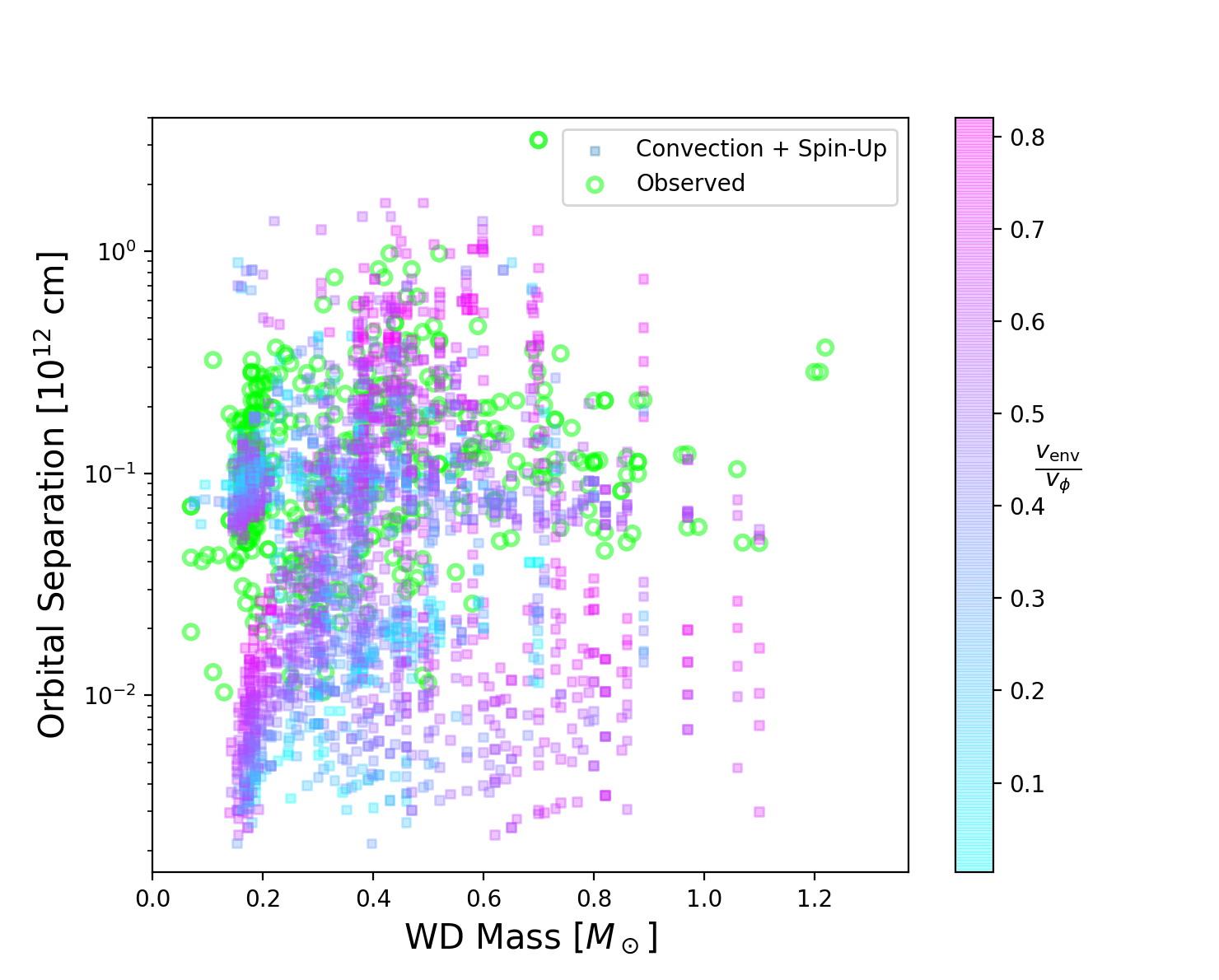}
    \caption{A comparison of observed DWD orbital separations and final separations of modelled CEs with convection and spin-up. The $v_{\rm env}/v_\phi$ values are shown in a color gradient from $v_{\rm env}/{v_\phi}=0$ (stationary envelope) in cyan to $v_{\rm env}/{v_\phi}=1$ (co-rotating envelope) in magenta. The modelled systems with convection and spin-up match observed systems. There is an over-representation of very short-period ($<10^{10}$ cm) systems which may be due to the lack of an IMF in this work.}
    \label{fig:colorfulvolcano}
\end{figure}

Our predicted DWD separations closely match those observed (see Figure~\ref{fig:colorfulvolcano}). The green circles represent known DWD systems and are the same as those in Figure~\ref{fig:volcano}. The colourful squares represent the DWD components and are coloured by the calculated spin-up value. The parameter space filled by observations is also filled by models. The over-representation of very short period (${\lesssim}10^{10}$ cm) systems may be due to the lack of sampling from an IMF. 

\subsection{Spin-Up and Mass Ratio}

\begin{figure}
	\includegraphics[width=\columnwidth]{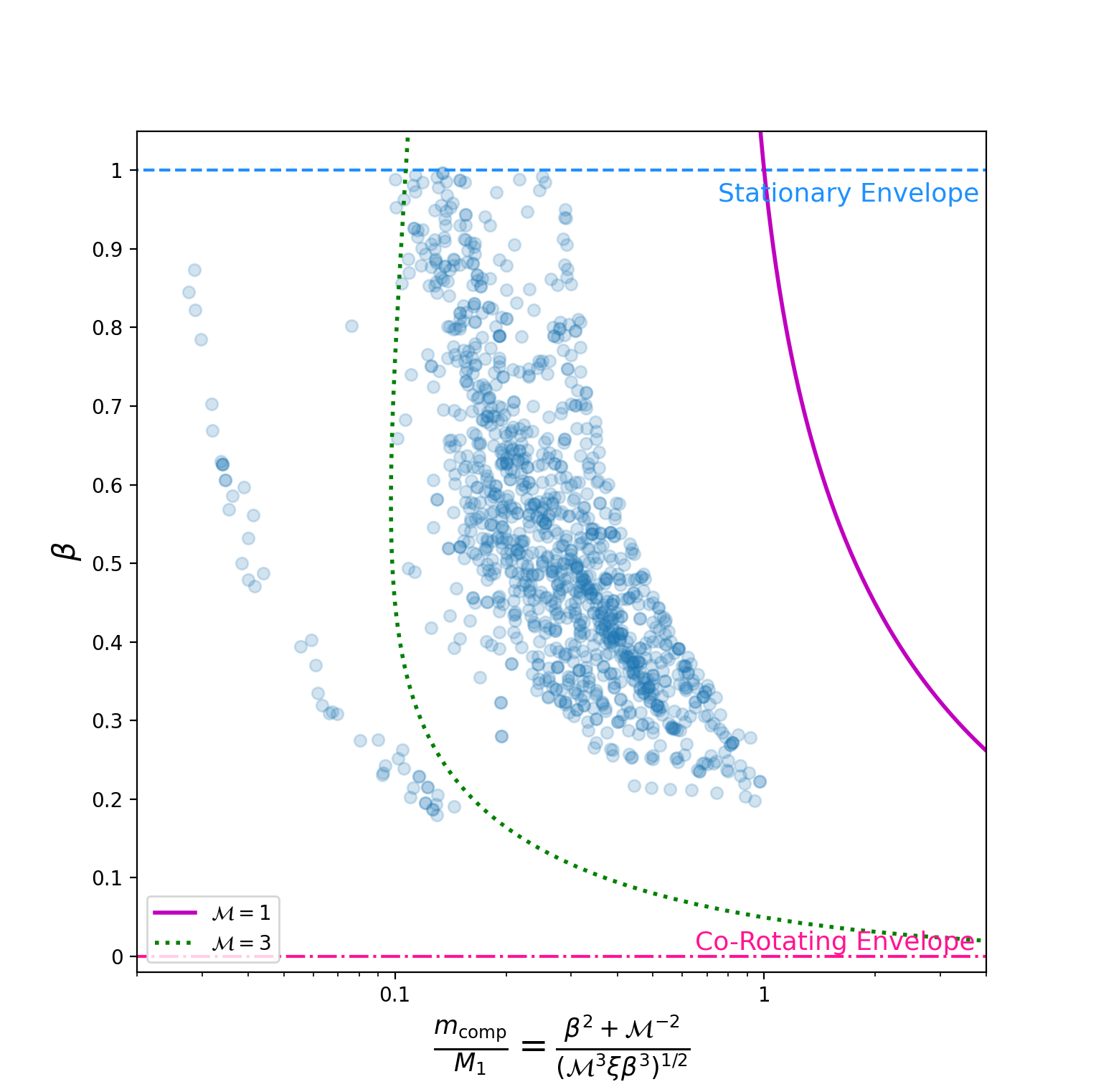}
    \caption{The $\beta$ value ($1-v_{\rm env}/v_\phi$) as a function of corresponding mass ratio of modelled systems. The mass ratio is calculated to always be less than unity and $\beta$ is calculated with use of model parameters as described in Equation~\ref{eq:beta4root}. The two curves follow $\beta$ as a function of the mass ratio and the Mach number, $\mathcal{M}$; $\mathcal{M}=1$ is shown in the solid magenta curve and $\mathcal{M}=3$ is shown in the dashed green curve. These curves bound the modelled data with few exceptions and $1 \le \mathcal{M} \le 3$ accurately describes the majority of the stellar interior models used.}
    \label{fig:qvsbeta}
\end{figure}
\begin{figure}
    \centering
    \includegraphics[width=\columnwidth]{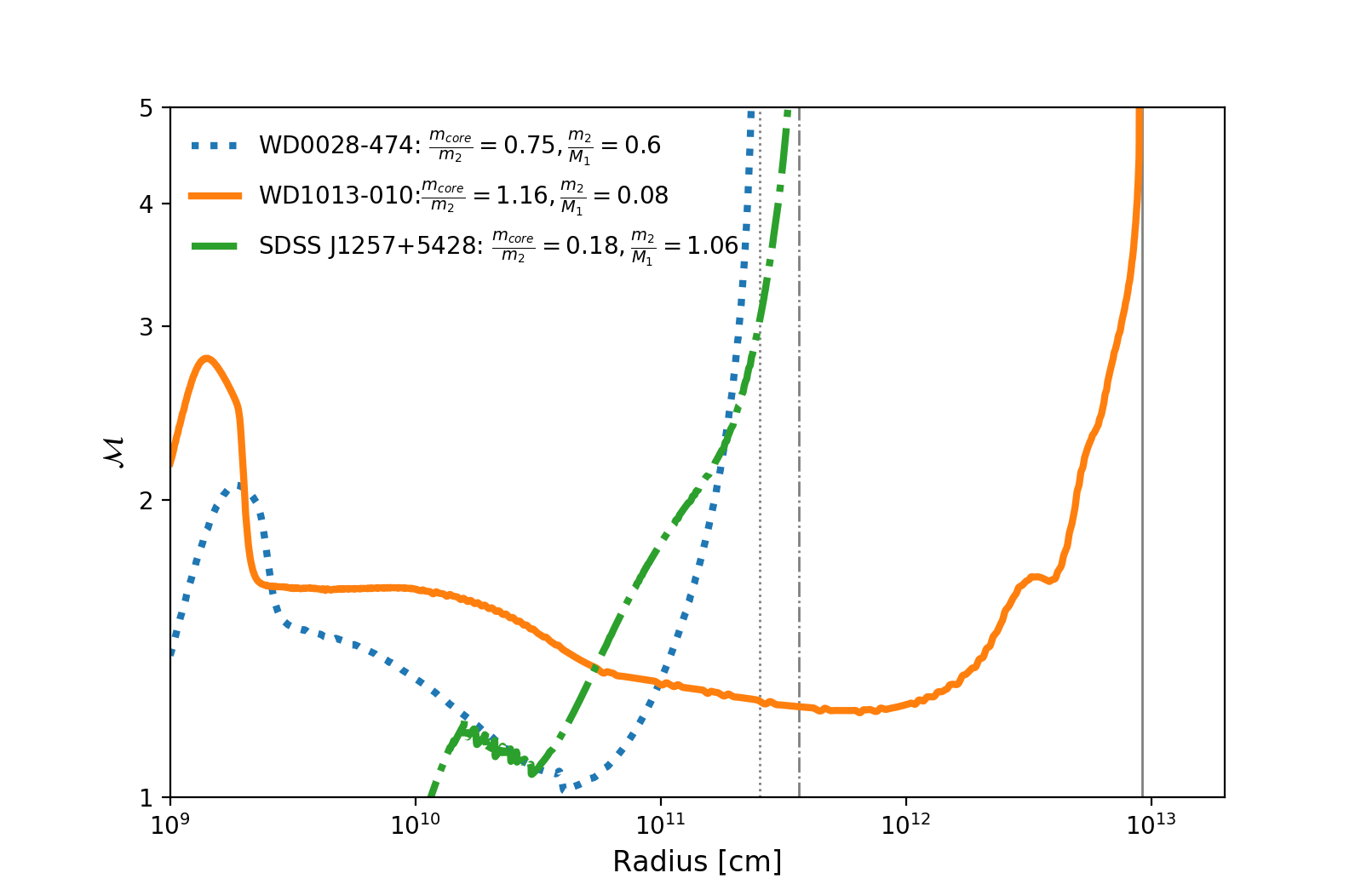}
    \caption{Mach number versus radius for three representative models. The dashed, blue curve follows the Mach number through a primary star of mass $1.0 M_\odot$ with core mass $0.45 M_\odot$ and companion of mass $0.6 M_\odot$, closely representing the WD0028-474 system. The solid, orange curve follows the Mach number through a primary star of mass $5.0 M_\odot$ with core mass $0.44 M_\odot$ and companion of mass $0.38 M_\odot$, closely representing the WD1013-010 system. The dot-dashed, green curve follows the Mach number through a primary star of mass $1.0 M_\odot$ with core mass $0.2 M_\odot$ and companion white dwarf of mass $1.06 M_\odot$, closely representing the SDSS J1257+5428 system. The gray vertical lines mark the outer radius of each primary star. The Mach values fall in the range $1<\mathcal{M}<3$ for all mass ratios in this work except where they asymptotically increase at the surface of the star.}
    \label{fig:Mach}
\end{figure}

There is a correlation between the mass ratios of the initial systems, $m_{\rm comp}/M_1$, and the amount of spin-up necessary such that the maximum convective luminosity is greater than the drag luminosity. As the mass ratio increases, the $\beta$ value decreases, i.e., the amount of spin-up necessary to meet the luminosity inequality increases (since $v_{\rm env}/v_\phi=1-\beta$). This relationship is shown with blue points in Figure~\ref{fig:qvsbeta}. 

The mass ratio of the initial system can also be mathematically related to $\beta$ by making the approximation $v_\phi \simeq c_s$, a reasonable assumption given that the Mach number, $\mathcal{M}$, is of order unity in these systems except near the stellar surface.  For three primary models, $\mathcal{M}$~vs.~r is shown in Figure~\ref{fig:Mach}. When the Keplerian velocity and the sound speed are set equal to each other in Equation~\ref{eq:beta4root}, the solution for the Keplerian velocity is
\begin{equation}
    \label{eq:vphi_simple}
    v_\phi^2 = \pm \frac{(\mathcal{M}^3 \xi)^{1/2} G m_{\rm comp} \beta^{3/2}}{r\left(\beta^2+\mathcal{M}^{-2}\right)}. 
\end{equation}

Since the Keplerian velocity is a function of $M_1$, this equality can be expressed in terms of the mass ratio ($m_{\rm comp}/M_1$), $\beta$, and $\mathcal{M}$ in the following way:
\begin{equation}
\label{eq:qvsbeta}
    \frac{m_{\rm comp}}{M_1}=\frac{\beta^2+\mathcal{M}^{-2}}{(\mathcal{M}^3 \xi \beta^3)^{1/2}}.
\end{equation}
The vast majority of $\beta$ values are bounded by the above equation evaluated at $\mathcal{M}=1$ and $\mathcal{M}=3$; these values are representative of the upper and lower limits of the Mach numbers within the primary's envelope. A plot of the above function in comparison to the relationship $\beta$~vs.~$m_{\rm comp}/M_1$ is shown in Figure~\ref{fig:qvsbeta}.

\section{Conclusion}
\label{sec:conclusion}

We considered how the effects of convection and spin-up in common envelope evolution impact the formation of DWDs. For each observed DWD system, two corresponding CEs were modelled: (i.) the more-massive WD as the companion and the less-massive WD as the core mass, and (ii.) the less-massive WD as the companion and the more-massive WD as the core mass.  To study convective effects, we employ detailed stellar interior models to compare the convective transport timescale to the inspiral timescale and the drag luminosity to the maximum luminosity that can transport energy via subsonic convection. The stellar envelopes are spun-up such that convection can accommodate the energy as the orbit decays. Our major findings are as follows.

\begin{itemize}[leftmargin=20pt,labelindent=5pt,itemindent=-5pt]
    \item[--] The correlation between the convective boundary and observed DWD separations reinforces the connection between short-period binaries and convective properties of the primary described in \citet{Wilson2019}. (See Figure~\ref{fig:volcano}.)
    \item[--] Our physically motivated description of ejection efficiency which combines convective effects with spinning-up the convective region of the envelope produces final separations of modelled systems that match observations of DWDs. (See Figure~\ref{fig:colorfulvolcano}.) 
    \item[--] In order for convection to transport the energy released as the orbit decays, the envelope must be moderately spun-up. The $v_{\rm env}/v_\phi$ values range from 0.0 to 0.82, with a peak between 0.5 and 0.7; the envelope is never required to spin-up faster than 82\% of the Keplerian speed to transport the full amount of orbital energy released during inspiral.
\end{itemize}

There are several promising directions for extending this work. Our physically-motivated ejection efficiency could be included in population synthesis models.  In particular, it would be interesting to see how Figure~\ref{fig:colorfulvolcano} changes when the physics described in this work are incorporated into a binary population synthesis code with a proper IMF.  High-resolution, global simulations of common envelopes do not include convection and radiation, both necessary ingredients for the effects described herein.  Given that RGB/AGB stars possess deep and vigorous convective zones, future numerical work could be focused on incorporating these effects in self-consistent ways.

\section*{Acknowledgements}
ECW and JN acknowledge support from the following grants:
NASA~HST-AR-15044, NTID~SPDI15992, and NSF~AST-2009713. The authors thank Noam Soker, Warren Brown, Na'ama Hallakoun, Gabriel Guidarelli, Silvia Toonen, Carles Badenes, Ashley Ruiter, Luke Chamandy, Eric Blackman, and Adam Frank for stimulating discussions.

\textit{Data Availability:} No new observational data were generated from this research; data underlying this article are available in the articles and supplementary materials of the referenced papers. Stellar interior models were derived from Modules for Experiments in Stellar Astrophysics (\textsc{MESA}) which is available at http://mesa.sourceforge.net.




\bibliographystyle{mnras}
\bibliography{CE+DWD} 


\bsp	
\label{lastpage}
\end{document}